\documentclass[pra,twocolumn,amssymb,superscriptaddress,noshowpacs]{revtex4}
\usepackage{amsmath}
\usepackage{amsthm}
\usepackage{graphicx}
\usepackage[dvips]{epsfig}
\usepackage{bbm}
\usepackage{braket}
\usepackage[usenames,dvipsnames]{color}
\usepackage{wasysym}
\usepackage{bbold}
\usepackage{rotating}
\usepackage{natbib}

\begin{document}
\title{Disorder-assisted quantum transport in suboptimal decoherence regimes}

\author{Leonardo~Novo}
\affiliation{Physics of Information Group, Instituto de Telecomunica\c{c}\~oes, P-1049-001 Lisbon, Portugal}
\author{Masoud Mohseni}
\affiliation{Center for Excitonics, Research Laboratory of Electronics, Massachusetts Institute of Technology, Cambridge, MA 02139, USA}
\affiliation{Google Research, Venice, CA 90291}
\author{Yasser Omar}
\affiliation{Physics of Information Group, Instituto de Telecomunica\c{c}\~oes, P-1049-001 Lisbon, Portugal}
\affiliation{CEMAPRE, ISEG, Universidade de Lisboa, P-1200-781 Lisbon, Portugal}

\date{27th March 2015}

\begin{abstract}
We investigate quantum transport in binary tree structures and in hypercubes for the disordered Frenkel-exciton Hamiltonian under pure dephasing noise. We compute the energy transport efficiency as a function of disorder and dephasing rates. We demonstrate that dephasing improves transport efficiency not only in the disordered case, but also in the ordered one. The maximal transport efficiency is obtained when the dephasing timescale matches the hopping timescale, which represent new examples of the Goldilocks principle at the quantum scale. Remarkably, we find that in weak dephasing regimes, away from optimal levels of environmental fluctuations, the \emph{average} effect of increasing  disorder is to improve the transport efficiency until an optimal value for disorder is reached. Our results suggest that rational design of the site energies statistical distributions could lead to better performances in transport systems at nanoscale when their natural environments are far from the optimal dephasing regime.
\end{abstract}

\maketitle


\section{Introduction}

Recently, the study of exciton transport in complex networks has received a boost, mainly due to recent experiments showing the presence of long-lived quantum coherences in exciton transfer in certain natural light-harvesting complexes (LHC's) \cite{NLHS3,NLHS2,NLHS1, qbio}. These coherences are present even at ambient temperature where quantum effects are not expected due to the interaction with a highly noisy environment. Moreover, such systems exhibit a high energy transfer efficiency \cite{efficiency_LHS}, so it is a relevant question to understand how the coherent evolution induced by the structure and the environmental fluctuations causes such an efficient transport.

In trying to explain these high efficiencies, the idea of environment-assisted quantum transport (ENAQT) has been proposed \cite{enaqw,role_quantum_coherence, masoud, noise_assisted_plenio, filippo, plenio2013}. In the simplest approach, the exciton transport can be modelled by a tight-binding Hamiltonian with site energies disorder, losses and a trap, interacting with the environment. In such systems, it was found that there are regimes of noise in which transport efficiency is improved. The gist of the idea is the following: in a disordered system, transport is suppressed due to localization \cite{localisation} caused by scattering and destructive interferences inside the medium; however, the presence of environmental fluctuations damps such interferences and localization can be overcome. This is not the only case where transport can be enhanced by interaction with the environment. In Ref.~\cite{filippo}, it was found that either noise or disorder can, independently, increase transport efficiency in highly symmetric structures, due to the presence of large invariant subspaces which are not coupled to the trapping site. Moreover, in Ref.~\cite{ENAQT_ordered} it has also been found that in ordered systems the presence of dephasing noise can enhance transport efficiency. General considerations about the scaling of mean trapping time with dephasing, in the weak and strong dephasing regimes, are presented in Ref.~\cite{wu}. It is also worth mentioning that in the scenario where the system reaches a non equilibrium steady state, analytical results regarding the conductivity as a function of disorder and dephasing strengths are known for a chain \cite{marko}. In general, it is still an open question to understand what are the kinds of structures that lead to efficient transport in a noisy environment \cite{geometrical_effects}. The understanding of how the interplay between noise and disorder affect transport could have interesting applications, for example, in the design of devices for photodetection, bio-sensing or photovoltaic light-harvesting systems.

In this work, we give a step towards this goal by studying quantum transport on two structures: binary trees and hypercubes. Binary trees are interesting structures which appear in a variety of applications including quantum algorithms \cite{algorithm_trees1, algorithm_trees2} and as a possible structure for artificial light-harvesting systems \cite{artificial_LHS, dendrimer_LHS}. The hypercube has been studied, for example, in the context of quantum state transfer \cite{ST_hypercube} and of quantum search algorithms \cite{search_hypercube}. Our model of transport consists of a tight-binding Hamiltonian with losses and a trap, interacting with the environment via a Haken-Strobl (pure-dephasing) model \cite{haken_strobl}. We consider site energies disorder and analyse how the transport efficiency depends on the dephasing rate and on the amount of disorder. Our goal is to understand the full picture of the dynamical interplay between disorder and dephasing and its consequences for the transport efficiency.

 In the case of the binary tree, we start the quantum walk in a statistical mixture of the leaves and place the trap at the root of the tree, whereas in the case of the hypercube the quantum walk starts as a statical mixture of all the sites and the trap is placed at one of the vertices. We find that the regime which maximizes transport efficiency is when the timescale of dephasing matches the timescale of hopping, verifying the quantum Goldilocks principle \cite{energy_scales,quantum_goldilocks, efficient_estimation}. This principle was put forward as a design principle for efficient structures: it states that a convergence of typical timescales of the system and the environment leads to optimal transfer efficiency. In this optimal dephasing regime, the efficiency is very robust against disorder. However, below this regime of dephasing we find that disorder can be beneficial for quantum transport. In fact, there is an optimal value of disorder for a fixed dephasing rate, as long as the latter is below its optimal value. Remarkably, we observe that the qualitative behaviour of efficiency in the whole range of parameters of disorder and dephasing is very similar for both the hypercube and the binary tree. This is consistent with previous results \cite{filippo,wu} for structures with large symmetry, for which there is a large subspace of
the Hilbert space which is orthogonal to the trap. 

For a better and more intuitive understanding of the physics behind this transport problem, we plot the dynamics of the population at the trap and the coherences between the trap and neighbouring sites, for the transport in the binary tree. This way, we observe clearly that in the zero disorder and zero dephasing scenario, there is destructive interference at the trap leading to very low transport efficiency. We observe also that the addition of disorder and dephasing suppresses these destructive interferences, opening the path to the trap. As a result, the population at the trap as well as the coherences between the trap and neighboring sites lasts for longer times. It is particularly counterintuitive that although dephasing damps all coherences, the interplay between Hamiltonian dynamics and dephasing makes some
coherences last longer.

Finally, in the last section we focus on the possibility to use disorder as a tool to optimize transport efficiency. We study numerically how much improvement of the transport efficiency can be obtained from adding the optimal value of disorder, for different values of the dephasing rate. We see that the improvement is maximal in the weak dephasing regime and is washed out when dephasing reaches its optimal value. However, the improvement is quite significant in the suboptimal dephasing regime reaching values as large as 30\%. It should be noted that, in practice, one always has non-zero dephasing, but not necessarily optimal dephasing for quantum transport. This way, our results could be explored for engineering novel materials that exploit quantum interference effects in the presence of environmental fluctuations, namely, by engineering the distribution of site energies to achieve better performances.

The effect of disorder as a tool for optimization of transport efficiency has been also addressed in Refs.~\cite{Transport2011,buchleitner_pre}. In these works, the authors study the transport efficiency (defined in a different way than here) of a random configuration of chromophores inside a sphere. They conclude that configurations with high efficiency are \emph{very rare} and result from an optimization of the structure which leads to fast and coherent transport to the trap. However, a different conclusion was reached in Ref.~\cite{geometrical_effects}, which also studies random configurations of chromophores inside a sphere, but, among other differences, sets a lower bound for the distance between chromophores. The conclusion was that configurations that have high efficiency are \emph{not rare} as long as we have the right chromophore density. 

In contrast, in this work we study the behaviour of transport efficiency with the strength of random disorder
measured by the standard deviation of the site energies instead of trying to find a particular configuration of site-energies and couplings that optimizes efficiency. This scenario is thus easier to implement in experiments, if there is not a very accurate control over the site-energies of the system.
\section{Quantum transport with pure-dephasing noise}\label{sec:model}

Our system Hamiltonian describing the quantum transport of an excitation in a structure with N sites is given by the tight-binding model with nearest neighbour couplings:
\begin{equation}\label{eq:tight_binding}
H_S=\sum_{m=1}^N\epsilon_m\ket{m}\bra{m}+ \sum_{<m,n>}V_{mn}(\ket{m}\bra{n}+\ket{n}\bra{m}).
\end{equation}
The states $\ket{m}$ represent the wave function of an excitation localized at site $m$, $\epsilon_m$ is the energy it takes for an excitation to occupy site $m$ and $V_{mn}$ is the coupling between nearest neighbours and thus it is zero if sites $m$ and $n$ are not connected. 

We consider static disorder in the site energies $\epsilon_m$ by assigning random values from a normal distribution with mean $0$ and standard deviation $\delta_{\epsilon}$.
We will choose our units such that $\hbar=1$ and $V=1$, and so all values of energies and rates will be given in units of $V$.

We assume that the interaction of the system with the environment is dominated by white-noise captured within the Haken-Strobl model (pure-dephasing) \cite{haken_strobl}. This is a good approximation if we assume the system-bath interaction $H_{SB}$ causes rapid stochastic fluctuations of the site energies of the system, i.e.
\begin{equation}
H_{SB}=\sum_m q_m(t)\ket{m}\bra{m},
\end{equation}  
where the $q_m(t)$ are random Gaussian variables with two-point correlation functions given by
\begin{equation}
\left<q_m(t)q_n(0)\right>=\delta_{mn}\delta(t)\gamma_{\phi},
\end{equation} 
where $\gamma_{\phi}$ is a site-independent rate. This model assumes that the coupling to the environment and thus the standard deviation of the energy fluctuations is the same at each site. Furthermore, fluctuations at different sites are considered independent. If the time scale of the fluctuations is much smaller than the typical time scales of the system, the averaging over fluctuations results in the master equation of the Lindblad type \cite{lindblad,breuer}:
\begin{equation}
\dot{\rho}(t)=-i\left<\left[H_S+H_{SB}(t),\rho(t)\right]\right>=-i\left[H_S,\rho(t)\right]+L_{\phi}(\rho(t)),
\end{equation}
where $\rho(t)$ is the density matrix of the system and the Lindblad operator $L_{\phi}(\rho(t))$ is given by
\begin{equation}
L_{\phi}(\rho(t))= \gamma_{\phi}\sum_m\left[A_m\rho(t)A_m^{\dagger}-\frac{1}{2}A_m A_m^{\dagger}\rho(t)-\frac{1}{2}\rho(t)A_m A_m^{\dagger}\right],
\end{equation}
with generators $A_m=\ket{m}\bra{m}$. The dephasing term $L_{\phi}(\rho(t))$ damps all off-diagonal entries of the density matrix, suppressing superpositions of localized states at a rate $\gamma_{\phi}$, which is called the dephasing rate. Note that the pure-dephasing (Haken-Strobl) model is a simplified but useful model that has been successfully used in numerous studies in quantum optics, quantum information science, physical chemistry, and condensed matter physics. Its prediction becomes more realistic when the system is interacting with a thermal bath at high temperatures, where its effects can be modelled by white noise. Furthermore, we account for exciton loss at each site via the addition of an antihermitian term $H_{recomb}$ to the system Hamiltonian, with
\begin{equation}\label{eq:recombination}
H_{recomb}=-i \Gamma\sum_{m=1}^N\ket{m}\bra{m}
\end{equation}
where $\Gamma$ is the recombination rate. This term causes a damping of all entries of the density matrix.
Furthermore, we introduce a trapping term in one of the sites, the target site. This term reads
\begin{equation}\label{eq:trap}
H_{trap}=-i\kappa\ket{trap}\bra{trap},
\end{equation}
where $\kappa$ is the rate at which the exciton gets trapped.
With this, we can define the energy transport efficiency 
\begin{equation}\label{transporteff}
\eta=2\kappa\int_0^{\infty} dt \bra{trap} \rho(t)\ket{trap}
\end{equation}
as the probability of the exciton being trapped at the trapping site, instead of its energy being dissipated through recombination. This will be the figure of merit of our model. 

Finally, the master equation governing the evolution of the system is given by
\begin{equation}\label{eq:master_equation}
\dot{\rho}(t)=-i \left(H\rho-\rho H^{\dagger}\right)+ L_{\phi}(\rho(t)),
\end{equation}
where the total Hamiltonian $H$ results from adding the contributions of Eqs.~\eqref{eq:tight_binding}, \eqref{eq:recombination} and \eqref{eq:trap}:
\begin{equation}\label{eq:total_H}
H=H_S+H_{trap}+H_{recomb}.
\end{equation}
This model has been used in the studies of excitonic transport in several works, such as Refs.~\cite{masoud, noise_assisted_plenio, ENAQT_ordered}. \\
\section{Quantum transport in binary tree structures}\label{sec:trees}
Tree structures appear in quantum information in some algorithms designed as quantum walks \cite{algorithm_trees2,algorithm_trees1}. Furthermore, dendrimer-like structures have been proposed as artificial light harvesting systems \cite{artificial_LHS,dendrimer_LHS}. 
The nearest neighbour tight-binding Hamiltonian \eqref{eq:tight_binding} for a binary tree with $g$ generations is:
\begin{align}\label{eq:hamiltonian_tree}
H_{tree}=&\sum_{m=1}^{2^g-1}\epsilon_m\ket{m}\bra{m}+\notag\\&V\sum_{m=1}^{2^{g-1}-1}(\ket{m}\bra{2m}+\ket{m}\bra{2m+1}+\text{h.c.}).
\end{align}

In this section, we investigate the transport of an excitation in the binary tree of 5 generations under the effect of disorder and dephasing, since these are factors that will naturally be present in an experimental implementation of these systems. First, in Sec.~\ref{sec:dis_deph}, we look at the transport efficiency as a function of these factors. In Sec.~\ref{sec:population_dynamics}, we look at the time evolution of the density matrix elements of the population at the trap (see Sec.~\ref{sec:model}) and coherence with neighbouring sites, which will give us a better understanding of the results obtained in Sec.~\ref{sec:dis_deph}. 
\begin{figure}[t]
  \centering
  \includegraphics[scale=0.7,trim=0pt 20pt 0pt 20pt]{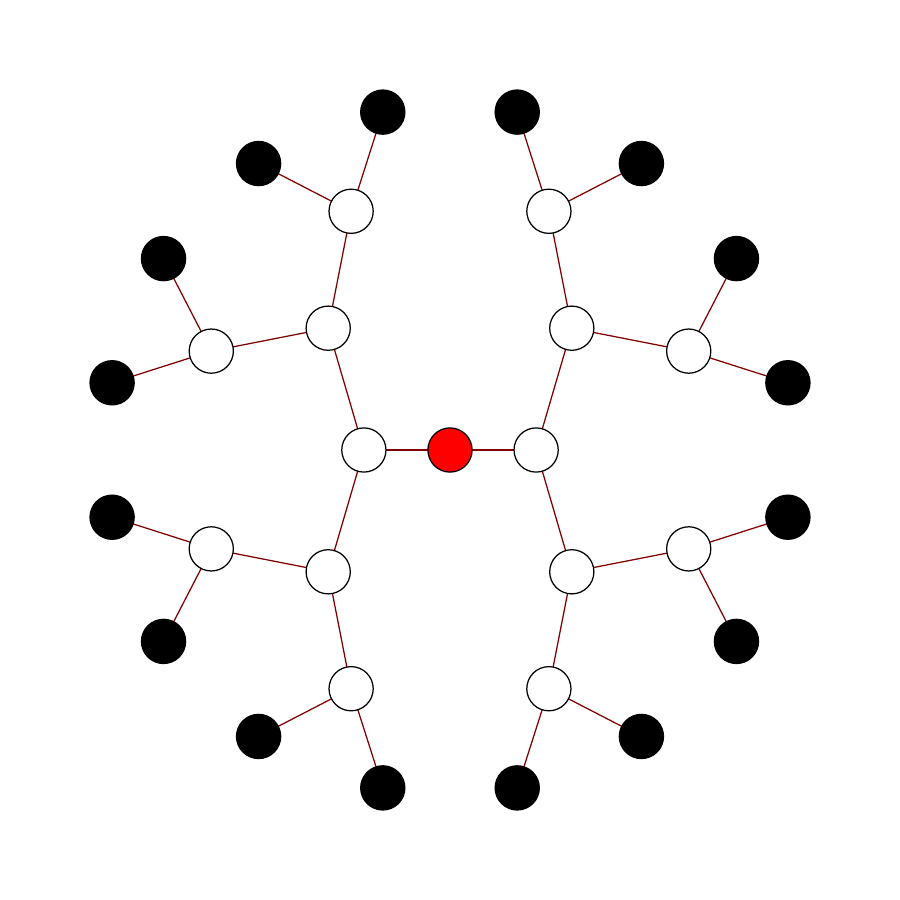}
  \caption{A binary tree structure with 5 generations. We set a statistical mixture of the leaves of the tree (black dots) as initial condition for the quantum transport and we place the trap at the root of the tree (red dot).}
  \label{fig:binary_tree_5}
\end{figure}
\subsection{Transport efficiency vs. disorder and dephasing}\label{sec:dis_deph}
\begin{figure}[h!]
\centering
\includegraphics[width=0.8\linewidth]{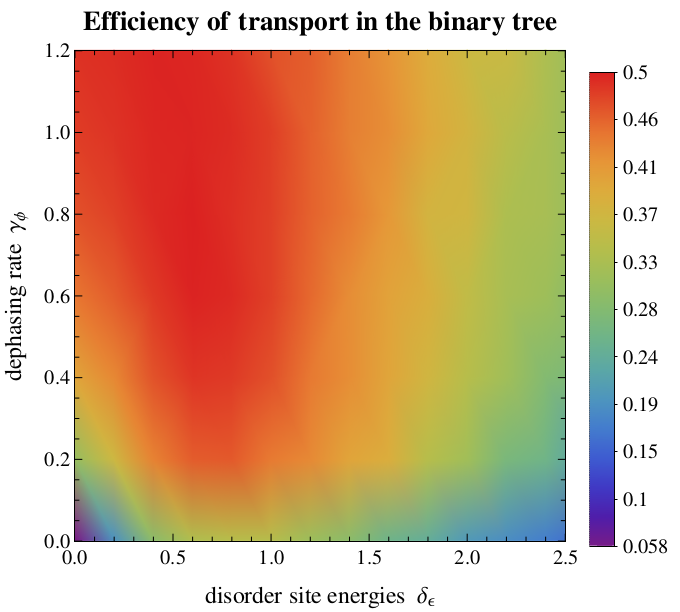}
\caption{Transport efficiency $\eta$ as a function of the disorder parameter $\delta_{\epsilon}$ and dephasing rate $\gamma_{\phi}$, in the binary tree with 5 generations. We consider as initial condition a statistical mixture at the leaves of the tree, and trapping at the root. The efficiency at each point is averaged over 100 random configurations for the site energies, where each site energy is sampled from a normal distribution with mean $0$ and standard deviation $\delta_{\epsilon}$. We observe a large optimality region around $\gamma_{\phi}\approx 1$. Below this region, we see that disorder improves significantly the transport efficiency until $\delta_{\epsilon}\approx 1$. All units are given in terms of the nearest neighbour coupling $V$.}
\label{plot:bin_tree_5}
\end{figure}

We study the quantum transport of an excitation in a binary tree of 5 generations, setting as initial condition an homogeneous statistical mixture of the leaves  and placing the trap at the root of the tree, site 1 (see Fig.~\ref{fig:binary_tree_5}). Furthermore, we set $\kappa=1$ and $\Gamma=0.01$, as typical trapping rate and dissipation energy scale, respectively, observed in biological light-harvesting complexes \cite{masoud}. We calculate numerically the efficiency of transport, defined in Eq.~\eqref{transporteff}, as a function of static disorder $\delta_{\epsilon}$ and of dephasing $\gamma_{\phi}$. We vary $\delta_{\epsilon}$ from $0$ to $2.5$ and $\gamma_{\phi}$ from $0$ to $1.2$. We recall that these values are always given in units of the nearest neighbours coupling $V$, as described in Sec.~\ref{sec:model}. The corresponding results are presented in Fig.~\ref{plot:bin_tree_5} where, at each point, the efficiency is averaged over 100 trees with random site energies of mean 0 and standard deviation $\delta_{\epsilon}$.

Analysing the results presented in Fig.~\ref{plot:bin_tree_5}, we start by noting that in the ordered case ($\delta_{\epsilon}=0$) with no interaction with the environment ($\gamma_{\phi}=0$) the transport efficiency is very low: $5.84\%$. In fact, it is the minimum efficiency in the range of parameters considered. Although at first sight this result might seem counter-intuitive, it can be explained by considering the invariant subspace of the Hamiltonian of the graph \cite{filippo}. The  invariant subspace is the subspace spanned by the eigenvectors of the Hamiltonian (in this case given by Eq.~\eqref{eq:hamiltonian_tree}) which have no overlap with the trap. The method to obtain this subspace is the following: if the eigenvector is not degenerate and has no overlap with the trap, it belongs to the invariant subspace; otherwise, in case there is a degenerate eigenspace of dimension $D$, one can always choose a basis such that $D-1$ eigenvectors of that subspace are not coupled to the trap, and thus also belong to the invariant subspace. If we set as initial condition of the transport problem a state which belongs to the invariant subspace, since it is an eigenstate and has no overlap with the trap, it will result in 0 transport efficiency, even if there are no losses through recombination. Thus, the only component of the initial state which can eventually be absorbed at the trapping site is the one which does not belong  to the invariant subspace. Let us call $D'$ the dimension of the invariant subspace and $\ket{\lambda_i}$ the vectors spanning it, with $i\in\{1,\dots,D'\}$. In the case of no disorder and no dephasing, and for a given initial condition $\rho_0$, we have:
\begin{equation}\label{eq:maximum_efficiency}
\eta\leq1-\sum_{i=1}^{D'}\bra{\lambda_i}\rho_0\ket{\lambda_i}.
\end{equation}
For the binary tree with 5 generations and setting the initial condition as a statistical mixture of the leaves of the tree
\begin{equation}
\rho_0=\sum_{i=16}^{31}\dfrac{1}{16}\ket{i}\bra{i},
\end{equation}
we obtain from Eq.~\eqref{eq:maximum_efficiency}
\begin{equation}\label{eq:upper_bound_eff_BT}
\eta\leq\dfrac{1}{16}.
\end{equation}
This inequality is saturated if there is no recombination ($\Gamma=0$). In Fig.~\ref{plot:bin_tree_5}, the efficiency without disorder and dephasing is $5.84\%$, only slightly smaller than the bound of $6.25\%$, because the recombination rate is small ($\Gamma=0.01$)  compared to the other energy scales of the system (couplings $V=1$, trapping rate $\kappa=1$).

\begin{figure}[h!]
\includegraphics[width=0.45\linewidth]{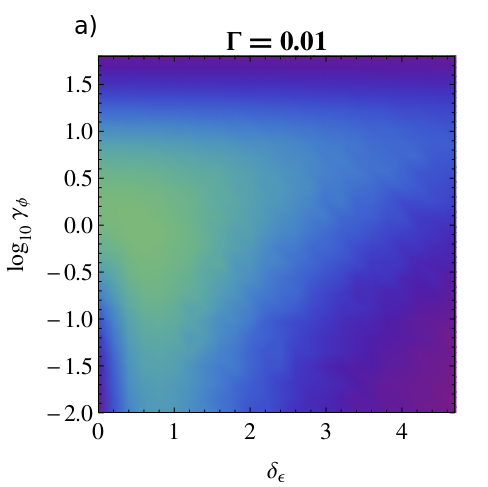}
\includegraphics[width=0.45\linewidth]{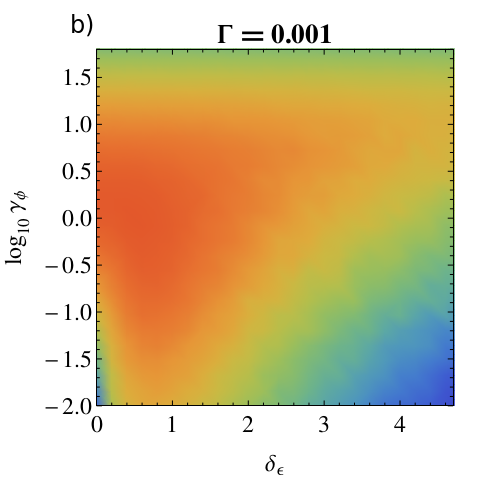}
\includegraphics[width=0.45\linewidth]{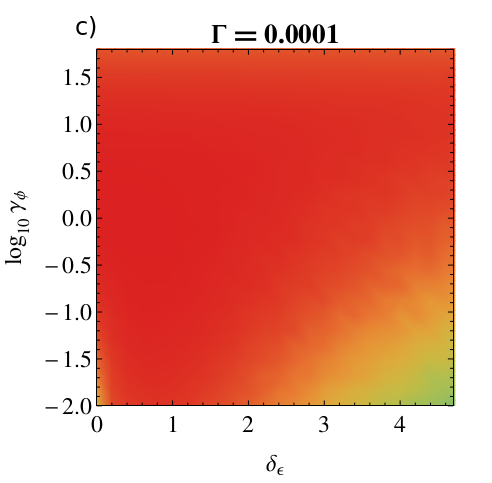}
\includegraphics[width=0.08\linewidth,trim = 0pt -18pt 0pt 0cm]{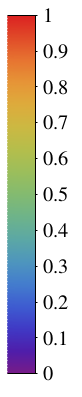}
\caption{Transport efficiency $\eta$ as a function of the disorder parameter $\delta_{\epsilon}$ and the logarithm of the dephasing rate $\gamma_{\phi}$, on the binary tree with 5 generations, for three orders of magnitude of the recombination rate: $\Gamma=10^{-2}, 10^{-3}, 10^{-4}$, plotted respectively in a), b) and c). We consider as initial condition a statistical mixture at the leaves of the tree and trapping at the root. The efficiency at each point is obtained by averaging over 100 random configurations of site energies disorder.  We verify that the optimality region increases with the inverse of the recombination rate. At very high dephasing rates, transport is suppressed due to Zeno effect. As the recombination rate decreases, the efficiency for 0 dephasing and 0 disorder tends to 1/16, as established in Eq.~\eqref{eq:upper_bound_eff_BT}, corresponding to the overlap of the initial condition with the eigenstates that couple to the trap. Outside the origin of the plot, the efficiency tends to 1 as $\Gamma$ tends to 0, because both disorder and dephasing destroy the invariant subspace. All units are given in terms of the nearest neighbours coupling $V$.}
\label{plot:variable_recombination}
\end{figure}

Still considering the ordered case ($\delta_{\epsilon}=0$), we observe that the efficiency increases with dephasing. This means that environment-assisted quantum transport takes place even in the ordered scenario. Thus, the usual justification of ENAQT, arguing that the presence of dephasing overcomes the localization due to disorder, cannot be applied here. However, this can be justified by the fact that the addition of dephasing will destroy, after some time, the invariant subspace and thus open new pathways through the network \cite{filippo}. Even if some dephasing can be good, though, too much dephasing will cause the freezing of transport due to quantum Zeno effect \cite{quantum_zeno}. So, there exists an optimal dephasing regime that maximizes transport efficiency. We find that the optimal value of dephasing is at $\gamma_{\phi}=1.6$. After this maximum, as we will see in Fig. \ref{plot:variable_recombination}, the efficiency starts decreasing and tends to zero in the limit of infinite dephasing, as expected due to the previous argument. Close to the optimal regime, the transport efficiency is very robust to disorder and remains almost constant as the latter varies within a large range of values (from $\delta_{\epsilon}=0$ to $\delta_{\epsilon}\approx1.5$). This is in agreement with the quantum Goldilocks principle \cite{energy_scales, quantum_goldilocks, efficient_estimation}, where the authors propose that the timescale associated with hopping and with dephasing should match for optimal and robust transport efficiency. 

Fig.~\ref{plot:bin_tree_5} also shows that for any given dephasing rate $\gamma_{\phi}$ below the optimal value, increasing the disorder from zero up to around $\delta_{\epsilon}\approx 1$ enhances transport efficiency. So, even though disorder is usually associated to the hindering of quantum transport, here we observe disorder-assisted quantum transport in the suboptimal dephasing regime. 
In the case of zero dephasing, this effect is explained in Ref.~\cite{filippo}: random site energies disorder also destroys the invariant subspace, making all eigenstates couple to the trap. However, for very small disorder, this coupling is weak and the transport to the trap slow. Thus, because of losses, the efficiency is still low. Very high values of disorder, though, can cause large energy mismatches between adjacent sites difficulting the propagation through the lattice. We see then that there must be an optimal disorder which maximizes transport efficiency.
Our results show that this optimal disorder also exists for each fixed dephasing rate in the suboptimal dephasing regime. This is consistent with the results of Ref.~\cite{wu}. The improvement due to disorder is maximal in the purely unitary case ($\gamma_{\phi}=0$), where the efficiency grows from $6\%$ (for $\delta_{\epsilon}=0$) to $34\%$ (for $\delta_{\epsilon}=0.83$). This improvement is washed out with increasing dephasing until the latter reaches its optimal value. This is understandable because dephasing destroys interferences, so high dephasing values mitigate the effect of disorder. However, for finite dephasing in the suboptimal regime, this effect can still be quite significant: for example, when the dephasing rate is 0.2, disorder improves the transport efficiency from $30\%$ (for $\delta_{\epsilon}=0$) to $47\%$ (for $\delta_{\epsilon}=0.8$).
Our results suggest that if such system, designed for quantum transport, has to work in the suboptimal dephasing regime, one could, in principle, engineer the site energies distribution in order to obtain better performances.

In order to understand the role of dissipation in the transport efficiency, we explore the effect of varying the recombination rate in Fig.~\ref{plot:variable_recombination}. In this figure, we show the dephasing axis in a logarithmic scale in order to observe a wider range of dephasing values and we vary the recombination rate three orders of magnitude: $\Gamma=10^{-2}, 10^{-3}, 10^{-4}$. For very high values of dephasing we observe that transport is suppressed due to the Zeno effect. As expected, for lower recombination rates there is an overall increase of the efficiency. In the case of $\Gamma=10^{-4}$ shown in Fig.~\ref{plot:variable_recombination} c), it is visible that the addition of disorder or dephasing destroys the invariant subspace: in the limit when $\Gamma$ tends to 0 the efficiency tends to 1 for any non-zero value of disorder or dephasing. In the case of zero disorder and zero dephasing (corresponding to the origin of the plot), the transport efficiencies are respectively: 5.84\% for $\Gamma=10^{-2}$, 6.20\% for $\Gamma=10^{-3}$ and 6.25\% for $\Gamma=10^{-4}$. Thus, as $\Gamma$ tends to 0, the efficiency tends to the value $\frac{1}{16}$ as predicted in Eq.~\eqref{eq:upper_bound_eff_BT}.

\subsection{Population and coherence dynamics at the trap}\label{sec:population_dynamics}

\begin{figure}
\includegraphics[width=0.48\textwidth]{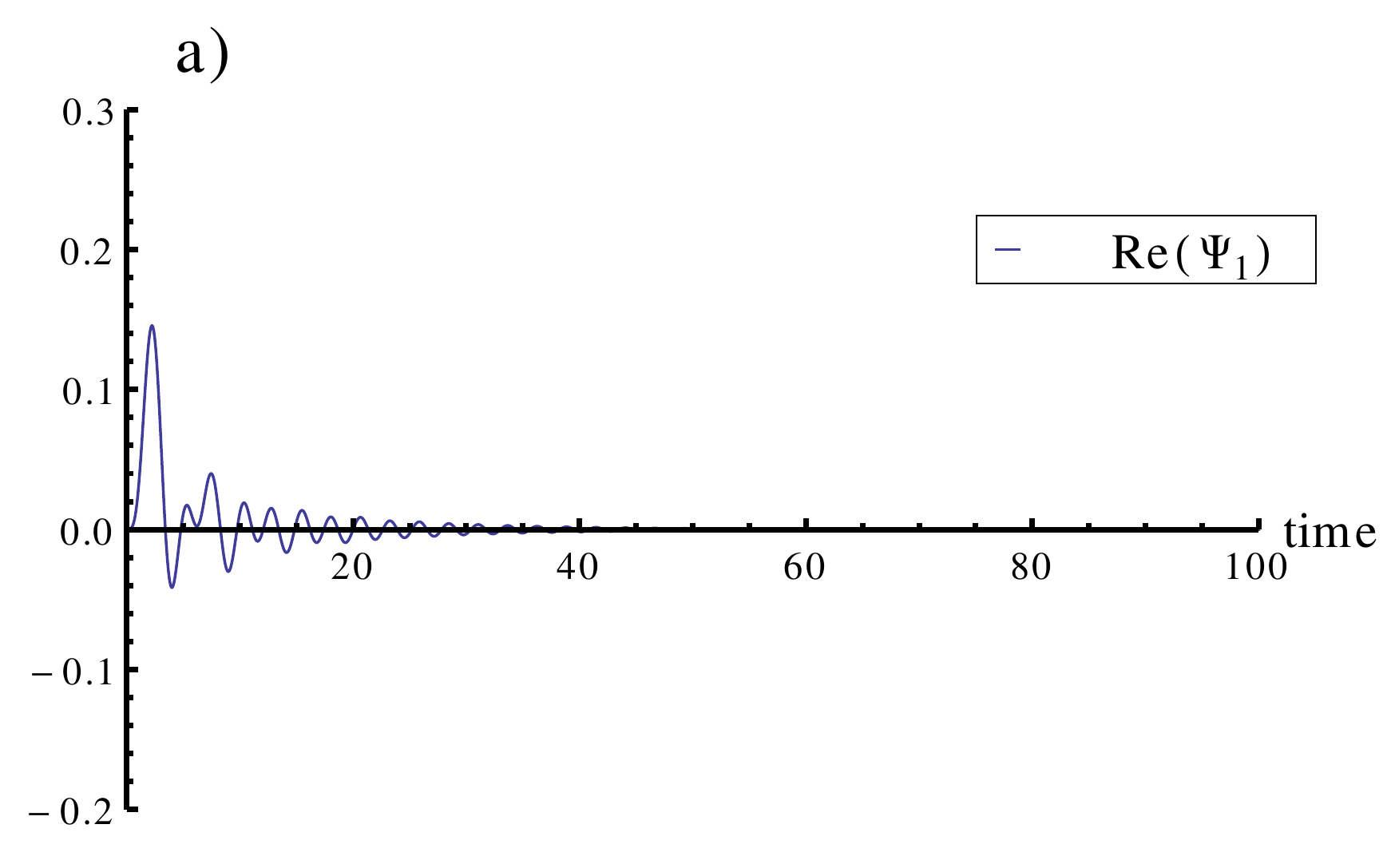}
\includegraphics[width=0.48\textwidth]{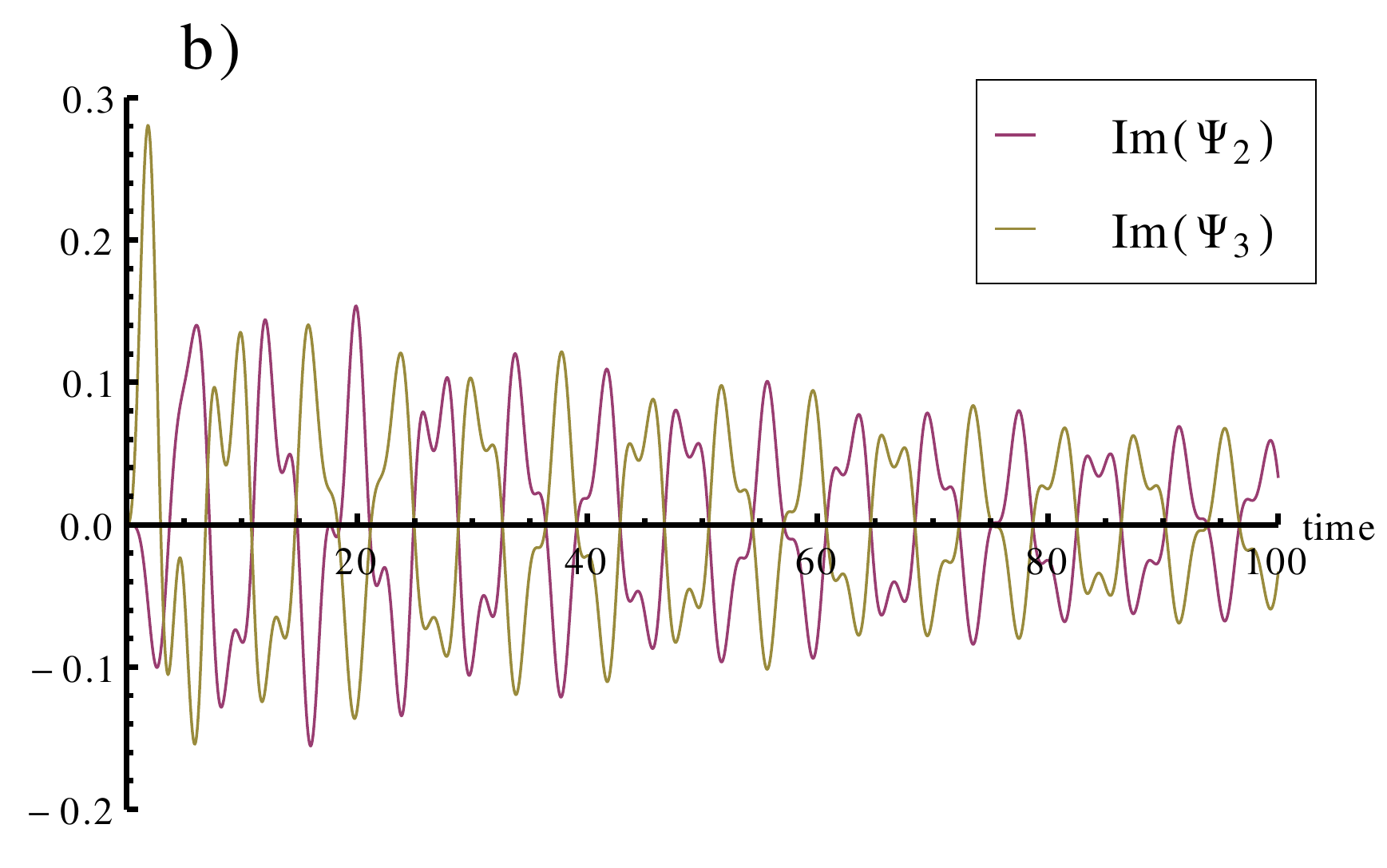}
\caption{Plots of: a) the real part of $\psi_1$; b) the imaginary part of $\psi_2$ and $\psi_3$. The imaginary part of $\psi_1$ is always 0, as is the real part of both $\psi_2$ and $\psi_3$. We choose as initial condition to have the exciton in a single leaf of the tree. Without loss of generality, we pick site 31, i.e.\ the initial state is $\ket{31}$. The time the excitation spends at the trap is much shorter than the time it spends outside the trap. This is due to the destructive interference between $\psi_2$ and $\psi_3$, as is clearly shown in b). We set the trapping rate $\kappa=1$ and the recombination rate $\Gamma=0.01$ as in Fig.~\ref{plot:bin_tree_5}. All units are given in terms of the couplings $V$.}\label{fig:WF}\end{figure}

In order to visualize better the dynamics of the exciton close to the trap when there is no disorder nor dephasing, we study the time-evolution of the wavefunction at the trap (site $1$) and adjacent sites ($2$ and $3$). Here, we set as initial condition the pure state $\ket{n}$. Writing the wave-function as $\ket{\Psi(t)}=\sum_{i=1}^n\psi_i(t)\ket{i}$, we obtain the evolution equation for $\psi_1(t)$:
\begin{equation}\label{eq:evolution_psi1}
\dot{\psi_1}(t)= -i(\psi_2(t)+\psi_3(t))-(\kappa+\Gamma)\psi_1(t),
\end{equation}
from the Schr\"{o}dinger equation with the Hamiltonian from Eq.~\eqref{eq:total_H}.
Thus, the evolution of $\psi_1$ is governed by the interference of $\psi_2$ and $\psi_3$ and damped at a rate $\kappa+\Gamma$. The solution of Eq.~\eqref{eq:evolution_psi1} is plotted in Fig.~\ref{fig:WF}. It is interesting to observe in this figure that the excitation lives much longer outside the trap than at the trap. After a short time, $\psi_2$ and $\psi_3$ synchronize (keeping opposite signs) and interfere destructively, preventing the excitation from reaching the trap. 
This helps us to understand why disorder and dephasing increase significantly the transport efficiency: both of them prevent this destructive interference from happening.

It is also interesting to observe the dynamics of the population at the trap $\rho_{11}$ and of the coherences between the trap and neighbouring sites $\rho_{12}$ and $\rho_{13}$, for different values of disorder and dephasing. We start again from the initial state $\ket{n}$. The evolution equation for $\rho_{11}$ is given by
\begin{equation}\label{eq:evolution_rho11}
\dfrac{\partial{\rho_{11}(t)}}{\partial t}=-2 \text{Im}(\rho_{12}(t)+\rho_{13}(t))-2(\kappa+\Gamma)\rho_{11}(t),
\end{equation} 
obtained from the master equation \eqref{eq:master_equation}. In Table~\ref{plots_rho11_rho12_rho13}, we present $\rho_{11}$, $\text{Im}(\rho_{12})$ and $\text{Im}(\rho_{13})$ as a function of time. The rows of the table correspond to three different dephasing rates (0, 0.2 and 1) and the columns to two values of disorder (0 and 1.4). The transport efficiency $\eta$, given by Eq.~(\ref{transporteff}), is proportional to the integral over time of $\rho_{11}$, so the longer the excitation remains at the trap, the higher the efficiency will be. This is precisely the effect of dephasing and disorder, as we can see in Table~\ref{plots_rho11_rho12_rho13}. Not only the excitation stays longer at the trap, but the coherence between sites $1$ and $3$ also lasts longer in the presence of dephasing and disorder. Despite the fact that dephasing alone damps off-diagonal terms of the density matrix, the interplay between dephasing and the driving Hamiltonian causes the coherences to last longer. Furthermore, we see clearly that in the suboptimal dephasing regimes the effect of adding disorder increases transport significantly whereas when dephasing is close to optimal, disorder does not help transport any longer. Indeed, the transport efficiency in the optimal dephasing region is very robust to disorder: changing disorder from 0 to $1.4$ only causes a few percent loss of efficiency. 
\section{Quantum transport in hypercubes}\label{sec:hypercube}

In this section, we study the transport efficiency as function of disorder and dephasing, for the quantum walk on the hypercube of dimension 4, similarly to what was done in Sec.~\ref{sec:dis_deph} for the binary tree. Here, we consider as initial condition a homogeneous statistical mixture of all sites, since all sites are equivalent in the graph. The hypercube is an interesting structure studied in different quantum information tasks, such as state transfer \cite{ST_hypercube} and quantum search \cite{search_hypercube}.

An hypercube of dimension $d$ has $2^d$ vertices. Each vertex can be labelled by a $d$-bit string and two vertices are connected if they differ by one bit. Thus, each vertex is connected to $d$ vertices and the adjacency matrix of the graph is given by:
\begin{equation}\label{adjacency_hypercube}
A=\sum_{j=1}^d\sigma^{(j)}_x,
\end{equation} 
where $\sigma^{(j)}_x$ represents the Pauli $\sigma_x$ matrix acting on bit $j$.
This way, the tight-binding Hamiltonian of the hypercube of dimension $d$ is given by:
\begin{equation}
H_S=\sum_{m=1}^{2^d}\epsilon_m\ket{m}\bra{m}+V\sum_{m,n}A_{mn}\ket{m}\bra{n}.
\end{equation}

\begin{figure}[h!]
\centering
\includegraphics[width=0.85\linewidth]{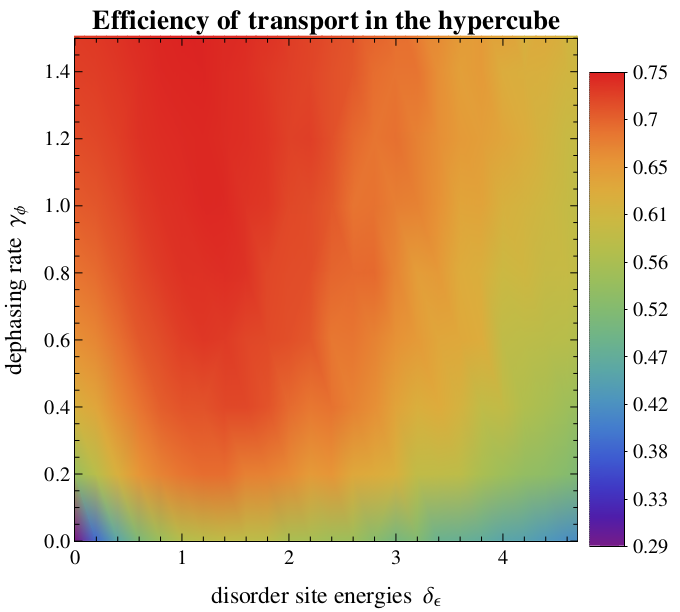}
\caption{Transport efficiency $\eta$ on the hypercube of dimension 4 as a function of the disorder parameter $\delta_{\epsilon}$ and dephasing rate $\gamma_{\phi}$, starting with a statistical mixture of all vertices and setting $\kappa=1$ and $\Gamma=0.01$. The efficiency at each point is obtained by averaging over 100 random configurations of site energies disorder. It is interesting to observe that the efficiency varies in a similar way as in the binary tree (see Fig.~\ref{plot:bin_tree_5}). Here, the optimality region is larger because we are dealing with a smaller structure. The disorder-assisted effect in suboptimal dephasing regimes is also visible in the hypercube. All units are given in terms of the nearest neighbour coupling $V$.}\label{plot:hypercube}
\end{figure}

In Fig.~\ref{plot:hypercube} and Fig.~\ref{plot:hypercube_variable_recomb} we present the variation of efficiency with respect to dephasing and disorder, as studied in the previous section for the binary tree (Figs.~\ref{plot:bin_tree_5} and \ref{plot:variable_recombination}) . Interestingly, the results obtained are qualitatively very similar for both structures. At zero disorder and zero dephasing, considering the concept of invariant subspaces (see Eq.~\eqref{eq:maximum_efficiency}), we obtain:
\begin{equation}
\eta\leq\dfrac{5}{16}.
\end{equation}
Once again, the disorder-assisted effect exists in the suboptimal dephasing regime and is washed out as the dephasing rate increases, disappearing when it reaches its optimal value of $\gamma_{\phi}\approx1$. To give an example, in the suboptimal dephasing regime with $\gamma_{\phi}=0.2$, the transport efficiency increases up to $16\%$ by introducing disorder, from $54\%$ (at $\delta_{\epsilon}=0$) to a maximum of $70\%$ (at $\delta_{\epsilon}=1.4$) . In fact, the main difference between both structures is that the values obtained for the efficiency are higher in the case of the hypercube than the ones obtained in Fig.~\ref{plot:bin_tree_5}, since we are now dealing with a smaller structure with only 16 nodes, in contrast with the binary tree structure considered before, which has 31 nodes.

\begin{figure}[h!]\label{plot:hypercube_variable_recomb}
\centering
\includegraphics[width=0.45\linewidth]{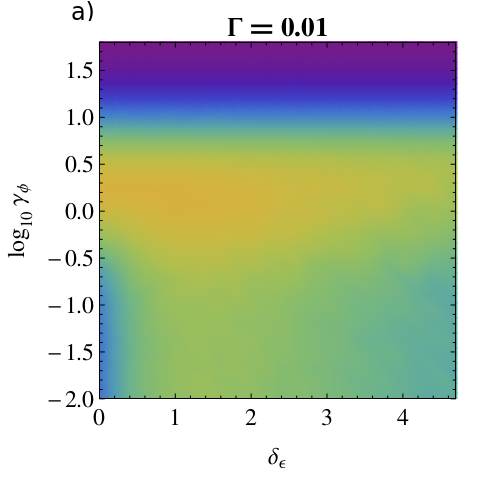}
\includegraphics[width=0.45\linewidth]{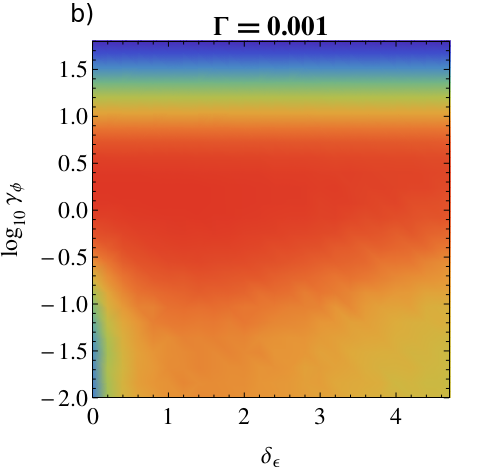}
\includegraphics[width=0.45\linewidth]{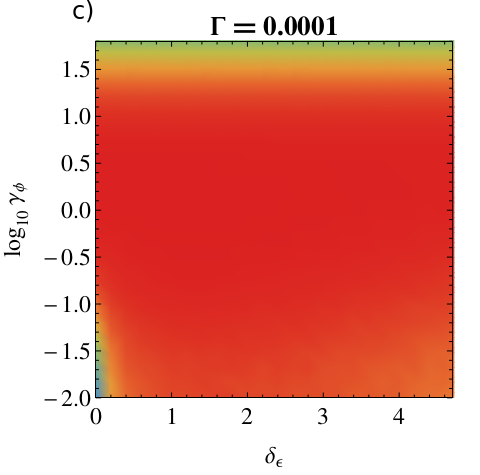}
\includegraphics[width=0.08\linewidth,trim = 0pt -18pt 0pt 0cm]{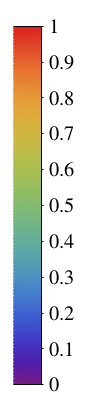}
\caption{\small{Transport efficiency $\eta$ on the hypercube of dimension 4 as a function of the disorder parameter $\delta_{\epsilon}$ and the logarithm of the dephasing rate $\gamma_{\phi}$, starting with a statistical mixture of all vertices with $\kappa=1$ and for different values of the recombination rate:  $\Gamma=10^{-2}, 10^{-3}, 10^{-4}$, plotted respectively in a), b) and c). The efficiency at each point is obtained by averaging over 100 random configurations of site energies disorder. The qualitative behaviour of the transport efficiency is very similar to the binary tree (see Fig.~\ref{plot:variable_recombination}). All units are given in terms of the nearest neighbour coupling $V$.}}
\label{plot:hypercube_variable_recomb}
\end{figure}
\section{Optimizing transport efficiency with disorder}
\begin{figure}
\centering
\includegraphics[width=0.9\linewidth,trim=0pt 0cm 0pt 0cm 0pt]{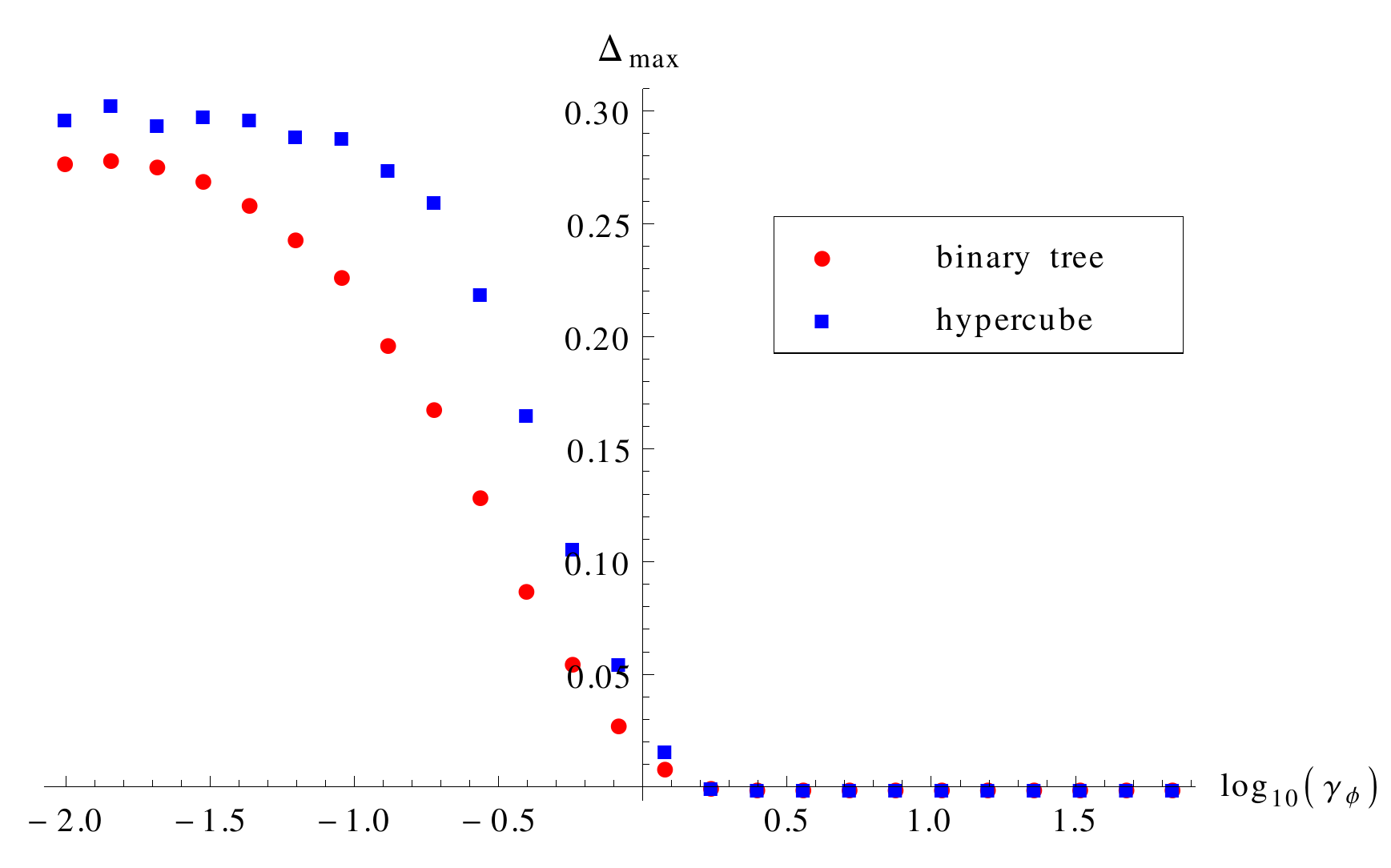}
\caption{Maximum improvement in the transport efficiency due to disorder, as defined in Eq.~\eqref{eq:delta}, here plotted as a function of the logarithm of the dephasing rate, for both the binary tree (5 generations) and the hypercube (dimension 4). The maximum improvement is approximately $30\%$, when the dephasing tends to 0. When dephasing reaches its optimal value of $\gamma_{\phi}\approx 1$ there is no more improvement due to disorder. The binary tree and the hypercube exhibit a similar behaviour.}
\label{plot:disorder_improvement}
\end{figure}

To understand better the significance of disorder in the improvement of transport efficiency, and for a fairer comparison between the effects of disorder in both structures, we define the \textit{maximum improvement of the transport efficiency due to disorder} $\Delta_{\text{max}}$, for a given dephasing rate $\gamma_{\phi}$, as the difference between the efficiency $\eta$ for the optimal value of the static disorder $\delta_{\epsilon}$ and the efficiency $\eta$ for $\delta_{\epsilon}=0$:
\begin{equation}\label{eq:delta}
\Delta_{max}(\gamma_{\phi})=\eta(\gamma_{\phi},\delta_{\epsilon}=\text{optimal})-\eta(\gamma_{\phi},\delta_{\epsilon}=0).
\end{equation}
This is the maximum amount by which efficiency can improve due to disorder, for a fixed dephasing rate. We plot this quantity in Fig.~\ref{plot:disorder_improvement} for the binary tree and the hypercube, using the data from Figs. \ref{plot:variable_recombination}.a) and \ref{plot:hypercube_variable_recomb}.a). The improvement due to disorder is maximal at zero dephasing. It decreases quickly as dephasing increases, reaching zero at the optimal dephasing regime. Once again, we observe that the qualitative behaviour is the same in both structures. It is clear from these results that disorder can indeed be used as a tool to significantly improve transport efficiency, if we are in the suboptimal dephasing regime. 

\pagebreak[3]
\begin{widetext}
~~~~~~~~~~~~~~~~~ ~~
\begin{table}[h!]
\centering
\begin{tabular}{ c| c| c| }
\cline{2-3}
 ~&\footnotesize{disorder $\delta_{\epsilon}=0$} &\footnotesize{disorder $\delta_{\epsilon}=1.4$} \\
\hline
\multicolumn{1}{ |c| }{\begin{sideways}\hspace{30pt}~\footnotesize{dephasing $\gamma_{\phi}=0$}\end{sideways}} & 
\includegraphics[width=0.47\textwidth,height=0.35\textwidth]{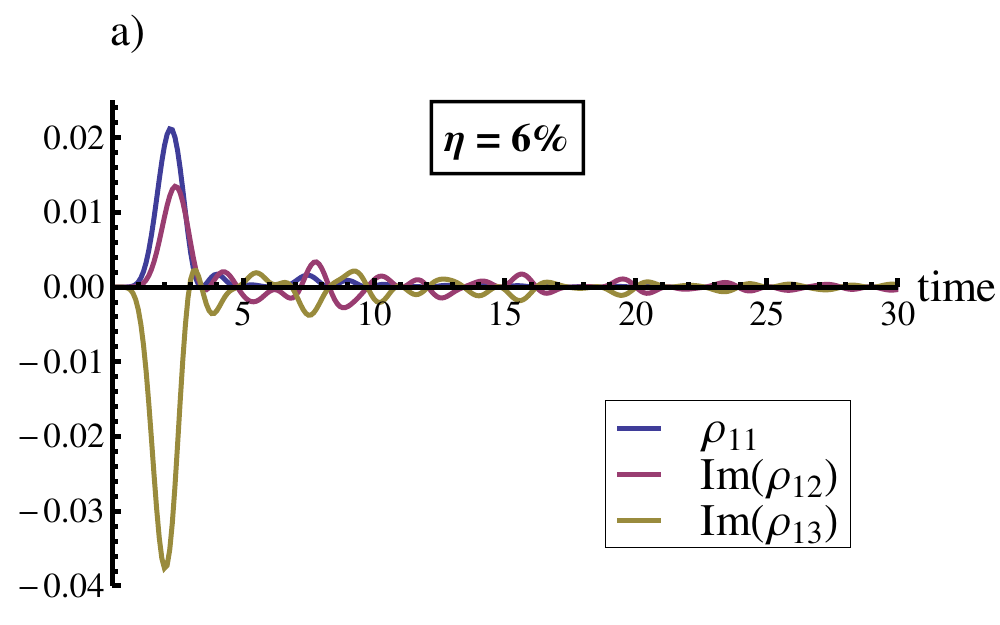} & 
\includegraphics[width=0.47\textwidth,height=0.35\textwidth]{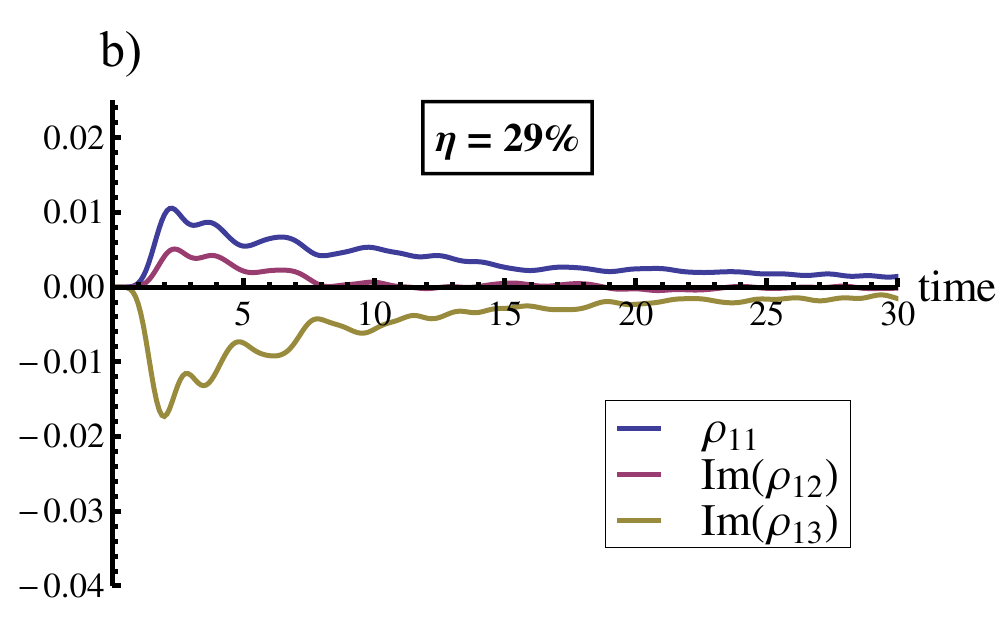}\\
 \hline
 \multicolumn{1}{ |c| }{\begin{sideways}\hspace{30pt}~\footnotesize{dephasing $\gamma_{\phi}=0.2$}\end{sideways}} & 
 \includegraphics[width=0.47\textwidth,height=0.35\textwidth]{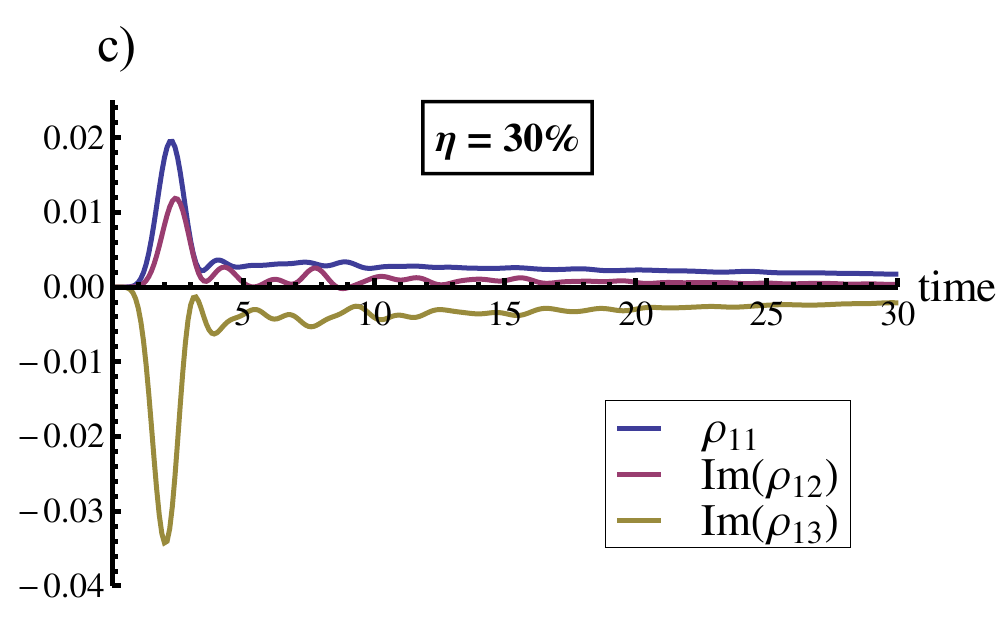} & 
 \includegraphics[width=0.47\textwidth,height=0.35\textwidth]{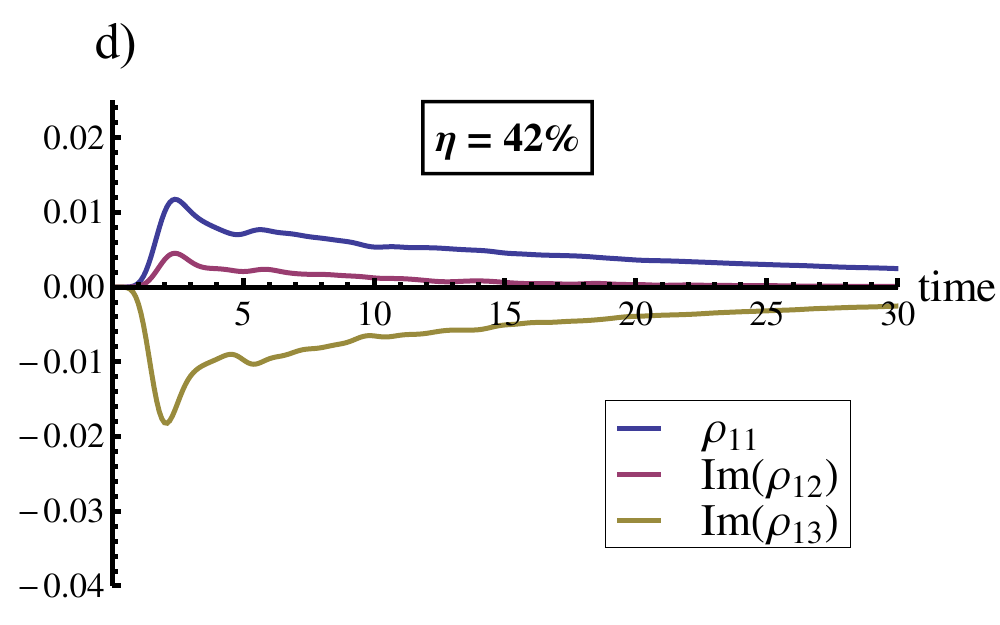} \\
 \hline
 \multicolumn{1}{ |c| }{\begin{sideways} \hspace{30pt}~\footnotesize{dephasing $\gamma_{\phi}=1$}\end{sideways}} & 
 \includegraphics[width=0.47\textwidth,height=0.35\textwidth]{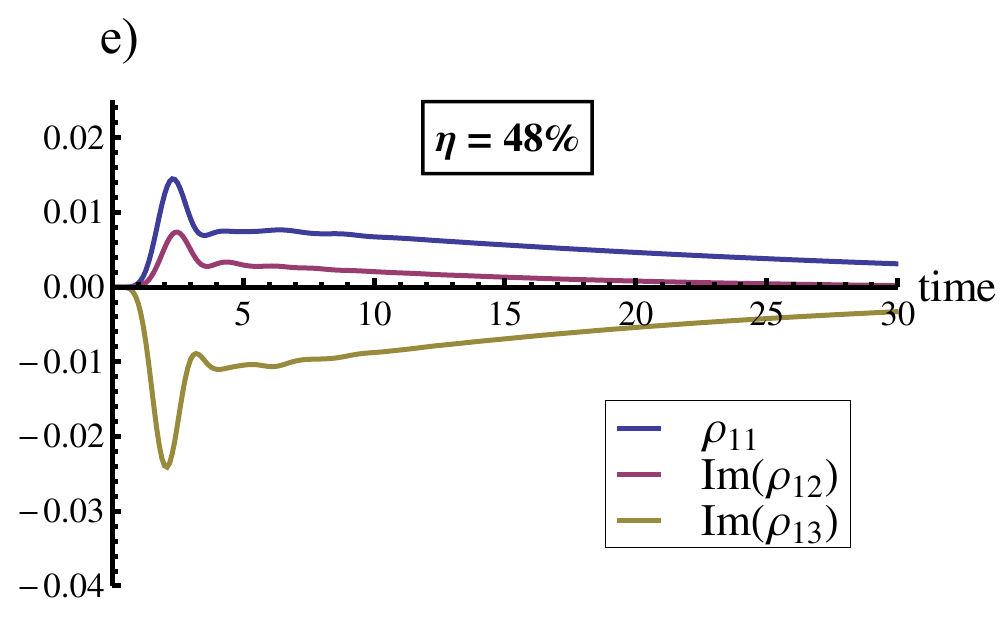} &
  \includegraphics[width=0.47\textwidth,height=0.35\textwidth]{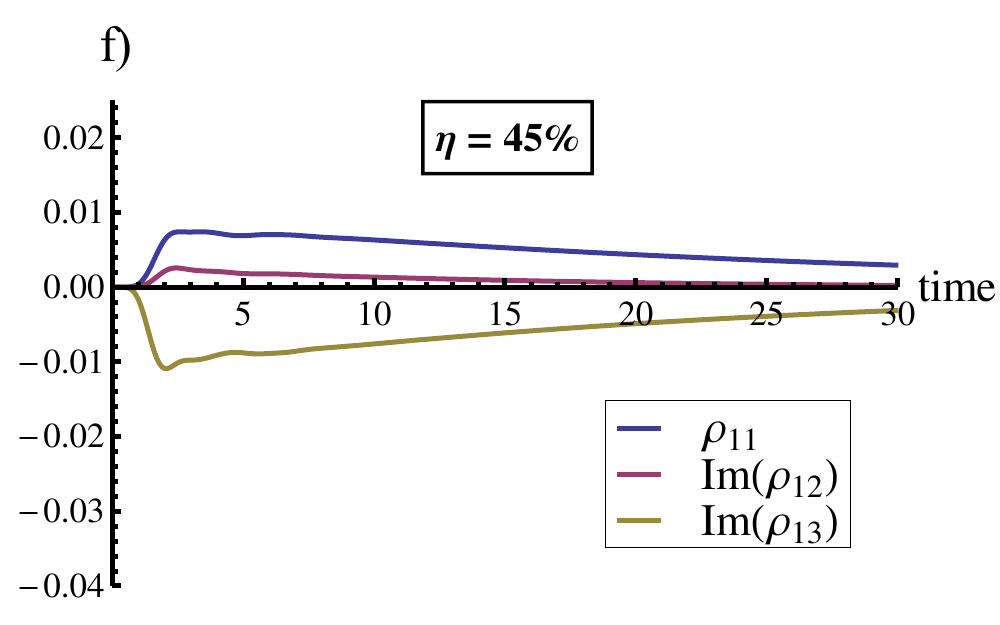} \\
\hline
\end{tabular}
\caption{In this table, we plot the population at the trap $\rho_{11}$ and the imaginary parts of $\rho_{12}$ and $\rho_{13}$, which are the terms coupled to $\rho_{11}$ in the master equation (see Eq.~\eqref{eq:evolution_rho11}). The rows correspond to different dephasing values (0, 0.2 and 1) and the columns to different disorder values (0 and 1.4). The plots with disorder $\delta_{\epsilon}=1.4$ are obtained after averaging over 100 random configurations of site energies. The initial condition is a pure state at the leaves of the tree (site $31$). It is clear that the population at the trap and superposition with neighbouring sites is prolonged with the addition of dephasing and disorder, which leads to an improvement in the transport efficiency. However, when dephasing is 1, the addition of disorder does not help transport any longer.}
\label{plots_rho11_rho12_rho13}
\end{table}
~~~~~~~~~~~~~~~~~~~~~~~~~~~~~~~~~~~~
\end{widetext}
\pagebreak
\section{Conclusions}
In this work we studied the efficiency of quantum transport in two different structures, the binary tree and the hypercube, as a function of site energies disorder and dephasing. We observed in these structures that, in the absence of disorder, the purely unitary evolution leads to low efficiency, and that the addition of dephasing and disorder can improve significantly the performance of transport. Optimality is reached when the dephasing rate matches approximately the hopping rate. At the optimal dephasing, the transport efficiency is very robust against disorder and it only decreases slightly if one increases disorder within a large range of values. However, when dephasing is below this optimal regime, disorder can improve transport efficiency in a very significant way, up to $30\%$ in our model. To better understand these results, we studied the wave function and the density operator population and coherences at the trapping site as a function of time. We observed that the addition of both disorder and dephasing can increase the time the excitation spends at the trap, contributing thus to an improvement of transport efficiency. We also found that the time the excitation remains delocalized between the trap and adjacent sites increases with both disorder and dephasing. It is remarkable that although dephasing alone damps superpositions between wavefunctions at different sites, its interplay with the hopping Hamiltonian leads to the prolongation of certain superpositions. Here, we do not deal with the issue of entanglement in single excitonic transport \cite{entanglement_LHC_plenio, sarovar_natphys}
 since we believe it  would not provide additional information beyond what is already being captured by quantum coherence.
 
These studies suggest that by engineering the distributions of disorder in excitonic transport systems, better performances can be achieved, if decoherence is below the optimal regime. This has potential application for the design of artificial light-harvesting systems or photosensing devices. 

\section*{Acknowledgements}
The authors thank Janos Asboth and Filippo Caruso for useful discussions. L.N. and Y.O. gratefully acknowledge the support from Funda\c{c}\~{a}o para a Ci\^{e}ncia e a Tecnologia (Portugal), namely through programmes PTDC/POPH and projects PEst-OE/EGE/UI0491/2013, PEst-OE/EEI/LA0008/2013, UID/EEA/50008/2013, IT/QuSim and CRUP-CPU/CQVibes, partially funded by EU FEDER, and from the EU FP7 projects LANDAUER (GA 318287) and PAPETS (GA 323901). Furthermore, LN acknowledges the support from the DP-PMI and FCT (Portugal) through scholarship SFRH/BD/52241/2013. M.M. thanks the support from QUBE and Google.

\bibliographystyle{apsrev}
\bibliography{biblio}

\end{document}